\documentclass[english,letterpaper,american,preprint,showpacs]{revtex4}
\usepackage[T1]{fontenc}
\usepackage[latin1]{inputenc}
\usepackage{graphicx}
\usepackage{amssymb}

\makeatletter

\usepackage{color}
\usepackage{amsmath}

\makeatletter

\newcommand{\tl}{\tilde}

\newcommand{\tsf}{\textsf{s}}

\newcommand{\txt}{\textnormal}
\newcommand{\eps}{\varepsilon}

\newcommand{\D}[1]{_{\textnormal{#1}}}

\newcommand{\emn}{\varepsilon_{\textnormal{mn}}}
\newcommand{\emx}{\varepsilon_{\textnormal{mx}}}
\newcommand{\cmx}{c_{\textnormal{{mx}}}}
\newcommand{\cmn}{c_{\textnormal{{mn}}}}
\newcommand{\Dmx}{\Delta_{\textnormal{mx}}}
\newcommand{\Dmn}{\Delta_{\textnormal{mn}}}

\usepackage{babel}
\makeatother
\begin{document}

\title{Stable Irregular Dynamics in Complex Neural Networks}

\author{Sven Jahnke$^{1-3}$, Raoul-Martin Memmesheimer$^{1-3}$ and Marc
Timme$^{1,2}$}

\affiliation{$^{1}$Network Dynamics Group, Max-Planck-Institute for Dynamics
\& Self-Organization (MPIDS), }

\affiliation{$^{2}$Bernstein Center for Computational Neuroscience (BCCN), 37073
G\"ottingen, Germany,}

\affiliation{$^{3}$Fakult\"at f\"ur Physik, Georg-August-Universit\"at G\"ottingen}

\date{Fri May 18 13:19:03 CEST 2006 }

\begin{abstract}
For infinitely large sparse networks of spiking neurons mean field
theory shows that a balanced state of highly irregular activity arises
under various conditions. Here we analytically investigate the microscopic
irregular dynamics in finite networks of arbitrary connectivity, keeping
track of all individual spike times. For delayed, purely inhibitory
interactions we demonstrate that the irregular dynamics is not chaotic
but rather stable and convergent towards periodic orbits. Moreover, every
generic periodic orbit of these dynamical systems is stable. These
results highlight that chaotic and stable dynamics are equally capable
of generating irregular activity.
\end{abstract}


\maketitle
Highly irregular dynamics is a prominent feature of multi-dimensional
complex systems and often attributed to chaos \cite{Baadi1999,Cvitanovic2005}.
Networks of spiking neurons, which interact by sending and receiving
electrical pulses (spikes) exhibit very irregular dynamics for a wide
range of conditions \cite{Vreeswijk1996,Brunel1999,Timme2002,Zumdieck2004,Hansel2001,Vreeswijk2000}.
For instance, networks of sparse random connectivity may display a
balanced state \cite{Vreeswijk1996} in which excitatory (positive)
and inhibitory (negative) inputs to each neuron balance on average
and only the fluctuations create spikes at irregular points in time
\cite{Vreeswijk1996,Brunel1999}. Their dynamics resemble random (e.g.~Poisson)
processes with low correlations across different neurons. Mean field
theory \cite{Brunel1999} shows that such irregular balanced activity
occurs in networks with excitatory and inhibitory recurrent feedback
as well as in networks that receive external excitatory inputs and
exhibit recurrent inhibition only \cite{Brunel1999,Timme2002}. 
Interestingly, for inhibitory networks recent work \cite{Jin2002,Zillmer2006}
suggests that this irregular dynamics might be stable, at least in
 globally coupled networks and in networks without delay. In particular,
Zillmer et al.~\cite{Zillmer2006} numerically measured a negative
maximal Lyapunov exponent of the irregular, seemingly chaotic spiking
dynamics. 

In this Letter, we analytically investigate the microscopic dynamics
in finite networks of spiking neurons with delayed inhibitory interactions
\cite{Delay}. For networks of arbitrary connectivity we show that
almost all spike sequences generated by the network -- trajectories
of the dynamical system -- are stable, i.e.~sufficiently small perturbations
of such trajectories do not grow with time.  In particular, every
generic periodic orbit in such inhibitory networks is stable. Moreover,
the analysis indicates that all attractors in such networks
are stable periodic orbits, which, however,
may only be reached after very long and irregular, but also stable
transients.

Consider $N$ neurons that interact on a directed network by sending
and receiving spikes. The subthreshold dynamics of the neurons' membrane
potentials $V_{i}(t)$ at time $t$ are given by\begin{equation}
\frac{d}{dt}V_{i}=f_{i}(V_{i})+\sum_{j=1}^{N}\sum_{k\in\mathbb{Z}}\eps_{i,j}\delta\left(t-t_{j,k}^{\tsf}-\tau_{i,j}\right)\label{eq:Veq}\end{equation}
where a smooth function $f_{i}$ specifies the internal dynamics,
$\eps_{i,j}\leq0$ is the inhibitory coupling strength and $\tau_{i,j}>0$
the delay time of a synaptic interaction from neuron $j$ to neuron
$i$, and $t_{j,k}^{\tsf}$ determines the time of the $k$th spike
sent by neuron $j$. Here $\delta(.)$ is the Dirac delta-function.
If a neuron $j$ reaches the threshold potential, $V_{j}(t^{-})=V_{\Theta,j}$,
it generates a spike at $t=:t_{j,k}^{\tsf}$ for some $k$ and is
reset, $V_{j}(t_{j,k}^{\tsf})=0$. We here require that the $f_{j}$
for all $j$ satisfy $f_{j}(V_{j})>0$ and $f_{j}{}^{\prime}(V_{j})<0$
for all $V_{j}\leq V_{\Theta,j}$ such that in isolation each neuron
exhibits oscillatory dynamics.

The neurons are equivalently described \cite{Mirollo1990} by a phase-like
variable $\phi_{j}(t)\in\left(-\infty,\phi_{\Theta,j}\right]$ satisfying
the linear differential equation \begin{equation}
d\phi_{j}/dt=1\label{eq:phidot}\end{equation}
at all non-event times. Upon reaching a phase threshold $\phi_{j}({t_{j,k}^{\tsf}}^{-})=\phi_{\Theta,j}$,
this phase is reset, $\phi_{j}(t_{j,k}^{\tsf}):=0$ and a spike is
generated. After a delay time $\tau_{i,j}$ that spike is received
by postsynaptic neuron $i$ of neuron $j$ and its phase changes according
to 

\begin{equation}
\phi_{i}(t_{j,k}^{\tsf}+\tau_{i,j})=H_{\eps_{i,j}}^{(i)}\left(\phi_{i}\left(\left(t_{j,k}^{\tsf}+\tau_{i,j}\right)^{-}\right)\right).\label{eq:phiupdate}\end{equation}
This interaction is mediated by the transfer function \begin{equation}
H_{\eps}^{(i)}(\phi)=U_{i}^{-1}\left[U_{i}(\phi)+\eps\right]\label{eq:H}\end{equation}
where $U_{i}(t)$ is the free (all $\eps_{i,j}=0$) solution
of (\ref{eq:Veq}) through the initial condition $U_{i}(0)=0$, yielding
$U_{i}'>0$ and $U_{i}''<0$, and $\phi_{\Theta,j}=U_{j}^{-1}(V_{\Theta,j})$,
cf.~\cite{Memmesheimer2006a}. For instance, for standard leaky integrate-and-fire
neurons, where $f_{i}(V)=I_{i}-\gamma_{i}V$ with time scale $\gamma_{i}^{-1}\geq0$
and equilibrium potential $\gamma_{i}^{-1}I_{i}>V_{\Theta,i}$, we
have $U_{i}(\phi)={\gamma_{i}^{-1}I}_{i}(1-\exp(-\gamma_{i}\phi))$.
Whereas the analysis below is valid for general $U_{i}(\phi)$, all
numerical simulations are presented for integrate-and-fire neurons.
In the following we consider arbitrary generic spike sequences in
which all neurons are active (i.e. there is a finite $T>0$, arbitrarily
large, such that in every time interval $[t,t+T)$, $t\in\mathbb{R}$,
every neuron fires at least once) and no two events occur at the same
time. Sending and receiving of spikes are the only nonlinear events
occurring in these systems.

\begin{figure}
\begin{centering}\includegraphics[clip,width=108mm,keepaspectratio]{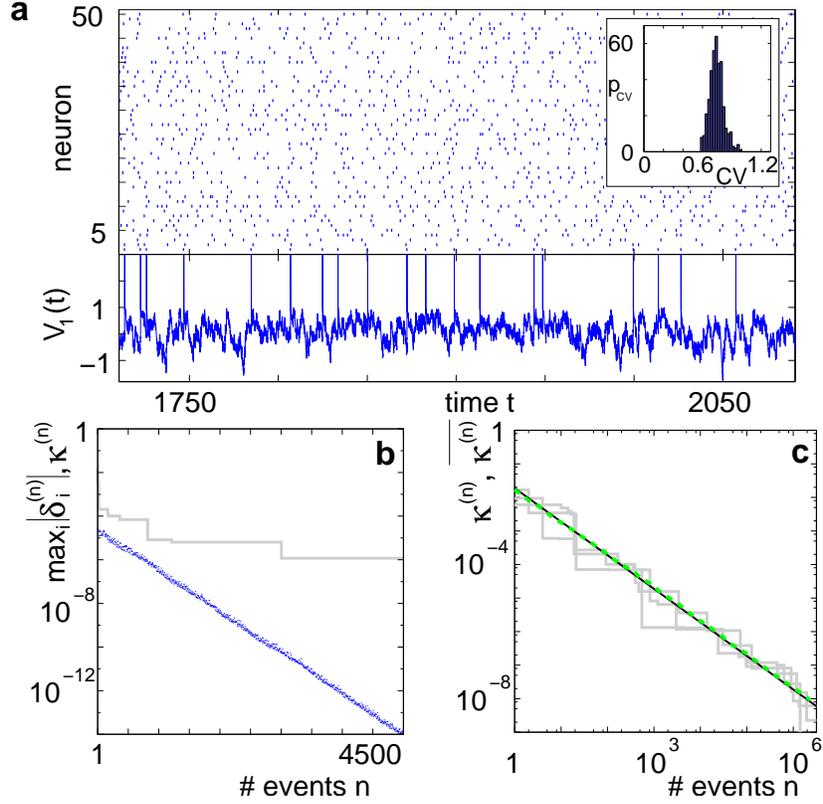}\par\end{centering}

\caption{(color). Stable irregular dynamics in a random network  ($N=400$,
$\gamma{}_{i}^{-1}I_{i}\equiv4.0$, $V_{\Theta,i}\equiv1$, $\tau_{i,j}\equiv0.1$,
connection probability $p=0.2$, $\sum_{j}\eps_{i,j}=-16$). (a) The
upper panel displays the spiking times (blue lines) of the first $50$
neurons. The lower panel displays the membrane potential trajectory
of neuron $i=1$ (spikes of height $\Delta U=2$ added at firing times).
The inset shows a histogram of the coefficients of variation $\textsf{CV}_{i}:=\sigma_{i}/\mu_{i}$;
$\mu_{i}=\left\langle t_{i,k+1}^{\tsf}-t_{i,k}^{\tsf}\right\rangle $;
$\sigma_{i}^{2}:=\left\langle \left(t_{i,k+1}^{\tsf}-t_{i,k}^{\tsf}-\mu_{i}\right)^{2}\right\rangle $
averaged over time. (b) Exponential decay of the maximal perturbation
$\max_{i}\left|\delta_{i}^{(n)}\right|$ (blue dots) and the minimal
margin $\kappa^{(n)}$ (gray line) for one given microscopic dynamics.
(c) Algebraic decay of the average minimal margin, $\overline{\kappa^{(n)}}$
(green dashed line, averaged over $250$ random initial conditions)
and the analytical prediction (no free fit parameter) of $\overline{\kappa^{(n)}}$
(black solid line). Additionally we show the minimal margin $\kappa^{(n)}$
for three exemplary initial conditions (gray lines), including that
of panel b.  \label{fig:balanced}}
\end{figure}

Highly irregular spiking sequences (cf.~Fig~\ref{fig:balanced}a)
constitute a typical form of activity in these networks, suggesting
that the underlying dynamics may be chaotic. However, as we show below for 
networks of arbitrary connectivity, this dynamics generically is stable. 
To show this, we first analytically study the exact microscopic dynamics
of an original trajectory, as defined by the (arbitrarily irregular)
sequence of events generated by the network, and a slight perturbation
to it that keeps the order of events as in the original.

The time of the $n$th event (sending and receiving) occurring in
the entire network is denoted by $t_{n}$ in the original sequence,
and by $\tilde{t}_{n}$ in the perturbed sequence. Here simultaneous reception
of the same spike at different neurons constitute one event. Analogously,
at a given time $t$, we denote the phases of neuron $i$ by $\phi_{i}(t)$
and $\tl\phi_{i}(t)$, respectively.  Let \begin{equation}
\Delta_{i}^{(n)}=\left(\phi_{i}(t_{n})-\tl\phi_{i}(\tl t_{n})\right)-\left(t_{n}-\tl t_{n}\right):=\delta_{i}^{(n)}-\delta t^{(n)}\label{eq:DeltaDef}\end{equation}
denote the difference of the phases of neuron $i$ between the two
sequences after the $n$th and before the $(n+1)$st event, corrected
for the time shift $\delta t^{(n)}=t_{n}-\tilde{t}_{n}$ between the
sequences. If at the $(n+1)$st event some neuron $j^{\ast}$ sends
a spike, the phase shifts \begin{equation}
\Delta_{i}^{(n+1)}=\Delta_{i}^{(n)}\label{eq:DeltaSend}\end{equation}
 of all neurons $i$ stay unchanged. Due to the linear phase dynamics
(\ref{eq:phidot}) between the spikes, $\Delta_{j^{\ast}}^{(n+1)}=\Delta_{j^{\ast}}^{(n)}=-\delta t^{(n+1)}$
also specifies the temporal shift of the $(n+1)$st event. At some
$t_{l}=t_{n+1}+\tau_{i,j^{\ast}}$ postsynaptic neuron $i$ receives
the spike sent by $j^{\ast}$.  The resulting phase shifts are given
by \begin{equation}
\begin{array}{lll}
\Delta_{i}^{(l)} & \!\!= & \!\! H_{\eps_{i,j^{\ast}}}^{(i)}\!\left(\phi_{i}\left({t{}_{l}}^{-}\right)\right)-H_{\eps_{i,j^{\ast}}}^{(i)}\!\left(\tilde{\phi_{i}}\left({\tl{t}{}_{l}}^{-}\right)\right)-\delta t^{(l)},\end{array}\label{eqn:Delta1}\end{equation}
where $\phi_{i}\left({t{}_{l}}^{-}\right)=\phi_{i}(t_{l-1})+t_{n+1}+\tau_{i,j^{\ast}}-t_{l-1}$
and $\tilde{\phi_{i}}\left({\tl{t}{}_{l}}^{-}\right)=\tl\phi_{i}(\tl t_{l-1})+\tl t_{n+1}+\tau_{i,j^{\ast}}-\tl t_{l-1}$
are the phases just before spike reception.  Using the identities
$\phi_{i}\left({t{}_{l}}^{-}\right)=\tilde{\phi_{i}}\left({\tl{t}{}_{l}}^{-}\right)+\Delta_{i}^{(l-1)}+\delta t^{(n+1)}$
and $\delta t^{(l)}=\delta t^{(n+1)}=-\Delta_{j^{\ast}}^{(n)}$, we
apply the mean value theorem in eq.~(\ref{eqn:Delta1}) and obtain
\begin{equation}
\Delta_{i}^{(l)}=c_{i}^{(l)}\cdot\Delta_{i}^{(l-1)}+(1-c_{i}^{(l)})\cdot\Delta_{j^{\ast}}^{(n)}\label{eq:DeltaAveraging}\end{equation}
 where $c_{i}^{(l)}$ is given by the derivative $c_{i}^{(l)}=dH_{\eps_{i,j^{\ast}}}^{(i)}(\phi)/d\phi$
for some $\phi$ between $\phi_{i}\left({t{}_{l}}^{-}\right)$ and 
$\tilde{\phi_{i}}\left({\tl{t}{}_{l}}^{-}\right)$.
If neuron $j^{\ast}$ is not connected to neuron $i$, $\eps_{i,j^{\ast}}=0$,
the function $H_{\eps_{i,j^{\ast}}}^{(i)}(\phi)=H_{0}^{(i)}(\phi)=\phi$
is the identity map, such that the phase shift stays unchanged, $\Delta_{i}^{(l)}=\Delta_{i}^{(l-1)}$;
indeed $c_{i}^{(l)}=dH_{0}^{(i)}(\phi)/d\phi=1$ is independent of
$\phi$. If neuron $j^{\ast}$ is connected to $i$ we find $c_{i}^{(l)}$
bounded by \begin{equation}
\cmn\!:=\!\mathop{\txt{\small{min}}}\limits _{\phi,k}\{(H_{\emn}^{(k)})^{\prime}(\phi)\}\!\leq\! c_{i}^{(l)}\!\leq\!\mathop{\txt{\small{max}}}\limits _{\phi,k}\{(H_{\emx}^{(k)})^{\prime}(\phi)\}\!=:\!\cmx\label{eqn:averaging}\end{equation}
where $\emx=\max\limits _{i,j:\eps_{i,j}\neq0}\left\{ \eps_{i,j}\right\} $
and $\emn=\min\limits _{i,j:\eps_{i,j}\neq0}\left\{ \eps_{i,j}\right\} $.
The phase is confined to some finite interval $\phi\in\left[\phi\D{mn},\phi\D{mx}\right]$
which depends on the network parameters. Given that  $dH_{\eps}^{(i)}(\phi)/d\phi=U_{i}^{\prime}\left(\phi\right)/U_{i}^{\prime}\left(U_{i}^{-1}\left[U_{i}(\phi)+\eps\right]\right)$
 and using the monotonicity $U_{i}^{\prime}>0$ and concavity $U_{i}^{\prime\prime}<0$, we find $\cmn>0$ and $\cmx<1$, independent of the sequence
and of the network realization (including its connectivity). Thus
the phase shift after receiving a spike is a weighted average of earlier
shifts. 

Consider that the perturbed sequence is created from the original
one by perturbing the phases of all neurons and the sending times
of all spikes sent but not received at time $t=t_{0}$. (We denote
the maximum of these perturbations by $\Dmx^{(0)}$ and the minimum
by $\Dmn^{(0)}$). After some finite time all spikes perturbed initially
arrived and all perturbations to newly generated spike sending times
stem from previous perturbations of some neurons' phases \cite{Ashwin2005}.
Given that the maximum phase shift cannot increase and the minimum
not decrease according to (\ref{eq:DeltaAveraging}) and (\ref{eqn:averaging}),
this implies Lyapunov stability of the dynamical trajectories.

 For strongly connected networks \cite{StronglyConnected}, more
involved graph theoretical arguments show that the initial perturbations
actually decay exponentially such that the considered trajectories
are asymptotically stable (cf.~\cite{Jahnke,Timme2002}). Briefly,
following the propagation of a phase perturbation of one specific
neuron $l_{0}$ across the entire network, shows that after a finite
number of $K:=2NM$ events all perturbations (of the phases of all
neurons and of the sending times of spikes sent but not yet received)
are bounded from above by \begin{equation}
\Dmx^{(K)}\leq c^{\ast}\Delta_{l_{0}}^{(0)}+(1-c^{\ast})\Dmx^{(0)}.\label{eq:DeltaUpperBound}\end{equation}
Here $c^{\ast}:=(1-\cmx)^{N}(\cmn)^{2NM}\leq(1-\cmx)(\cmx)$ such
that $3/4\leq(1-c^{\ast})<1$ and $M$ depends on the spike sequence
but is finite. Similarly we find a lower bound given by $\Dmn^{(K)}\geq c^{\ast}\Delta_{l_{0}}^{(0)}+(1-c^{\ast})\Dmn^{(0)}$.
The difference of the maximal and the minimal perturbation after $K$
events is therefore given by \begin{equation}
\delta_{i}^{(K)}\leq\Dmx^{(K)}-\Dmn^{(K)}\leq(1-c^{\ast})(\Dmx^{(0)}-\Dmn^{(0)}).\label{eqn:convergence}\end{equation}
This inequality (\ref{eqn:convergence}) implies that both sequences
converge exponentially fast against each other. Thus all sequences
considered are asymptotically stable for all strongly connected networks.

A main condition for stability of trajectories was that the order
of events stays the same in the perturbed and original trajectories.
For arbitrary non-degenerate spike sequences, there is a non-zero
perturbation size keeping the order unchanged in any \emph{finite}
time interval. However, the requirement of an unchanged event order yields
more and more conditions over time such that the allowed size of a perturbation
could decay more quickly with time than the actual perturbation. This
will be excluded if the temporal \emph{margin} $\mu^{(n)}$ between
two subsequent events stays larger than the dynamical perturbation
for \emph{infinite} time. Formally, after time $t_{n}$ denote the
$k$th potential future event time (of the original trajectory) that
would arise if there were no future interactions by $\theta_{n,k}$
, $k\in\mathbb{N}$, and the temporal \emph{margin} by $\mu^{(n)}:=\theta_{n,2}-\theta_{n,1}$.
  A sufficiently small perturbation, satisfying $\Dmx^{(n)}-\Dmn^{(n)}\leq\mu^{(n)}$,
cannot change the order of the $(n+1)$st event.

This directly implies that almost all \emph{periodic orbits} (all
those with non-degenerate event times $t_{n}$) consisting of a finite
number of $P$ events are stable because there is a minimal margin
\begin{equation}
\kappa^{(P)}:=\min\limits _{n\in\{1,\ldots,P\}}\mu^{(n)}\label{eq:defMinMargin}\end{equation}
 for every non-degenerate periodic pattern. 

To further analyze stability properties of \emph{irregular non-periodic
spike sequences}, we consider the minimal margin $\kappa^{(n)}$ over
the first $n$ events. Assuming that, along with the irregular dynamics, 
the temporal margins are generated by a Poisson point process with
rate $\nu$, the distribution function of margins is given by $P\left(\mu^{(n)}\leq\mu\right)=1-e^{-\nu\mu}$.
The probability that the minimal margin $\kappa^{(n)}$ after $n$
events is smaller or equal to $\mu$ is determined by the probabilities
that not all individual margins $\mu^{(n)}$ are larger than $\mu$
such that \begin{equation}
P(\kappa^{(n)}\leq\mu)=1-\prod_{m=1}^{n}P(\mu^{(n)}>\mu)=1-e^{-n\nu\mu}\label{eq:minMarginDistr}\end{equation}
with density $\rho_{n}(\mu):=dP(\kappa^{(n)}\leq\mu)/d\mu=n\nu\exp(-n\nu\mu).$
This implies an algebraic decay with the number $n$ of events for
the expected minimal margin \begin{equation}
\overline{\kappa^{(n)}}=\int_{0}^{\infty}\mu\rho_{n}(\mu)d\mu=(\nu\cdot n)^{-1}\label{eqn:powerlaw}\end{equation}
that depends only on the event rate and is independent of the specific
network parameters. Numerical simulations confirm this \emph{algebraic}
decay (\ref{eqn:powerlaw}) of the expected minimal margin with the
number of network events $n$; an example is shown in Fig.~\ref{fig:balanced}c.
Together with the \emph{exponential} decay of dynamical perturbations
(\ref{eqn:convergence}) this indicates that for a sufficiently small
perturbation the order of events stays unchanged for \textit{infinite}
time. 

\begin{figure}
\begin{centering}\includegraphics[clip,width=108mm,keepaspectratio]{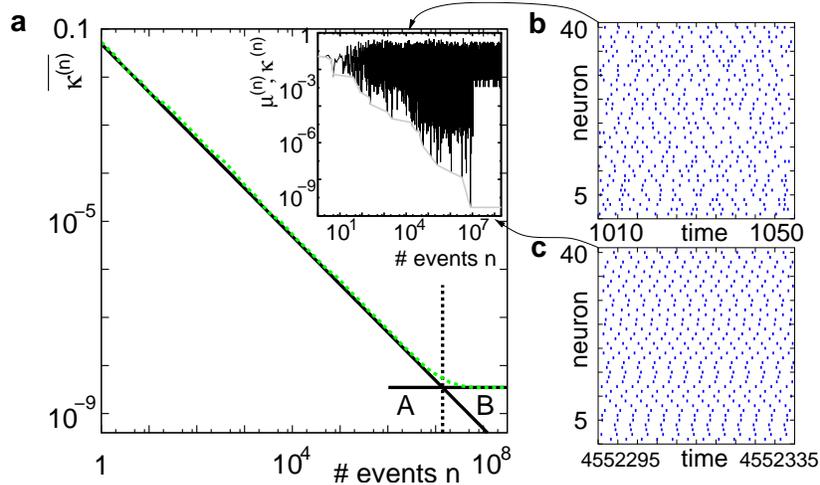}\par\end{centering}

\caption{(color) Convergence towards a periodic orbit in a random network
($N=40$, ${\gamma_{i}^{-1}I}_{i}\equiv3.0$, $V_{\Theta,i}\equiv1.0$,
$\tau_{i,j}\equiv0.1$, $p=0.2$, $\sum_{j}\epsilon_{i,j}=-3.3$).
(a) The average minimal margin $\overline{\kappa^{(n)}}$ (as in Fig.~\ref{fig:balanced}c)
decays as a power-law (region $A$) and saturates after about $10^{7}$
events (region $B$) when the periodic orbit is reached. Inset: Margin
$\mu^{(n)}$ (black) and minimal margin $\kappa^{(n)}$ (gray) for
a trajectory started from one specific initial condition. The margin
$\mu^{(n)}$ fluctuates strongly on the transient but is comparatively
large and bounded after the sequence becomes periodic; thus the minimal
margin $\kappa^{(n)}$ does not decrease further for future events
$n$. (b), (c): Snapshots of irregular spike sequences (b) after $n=20.000$
events on the transient and (c) after $n=10^{8}$ events on the periodic
orbit. \label{fig:periodic}}
\end{figure}

Interestingly, arbitrary irregular spike sequences necessarily converge
to a periodic orbit after finite time (cf.~Fig.~\ref{fig:periodic})
because there is some finite number $E$ such that (i) two sequences that share the same order of $E$ events are equally ordered for all future events 
because any
initial difference decays exponentially (\ref{eqn:convergence}) and
(ii) there is only a finite number of orderings of events in a finite
network such that a given sequence of length $E$ repeats after
finite time. Nevertheless, we find that the transient time until a
periodic orbit is reached, rapidly increases with network size $N$
and with the interaction strengths, in agreement with observations
in ref.~\cite{Zillmer2006}; cf.~also \cite{Jahnke}.   

 In summary, we analytically accessed the microscopic dynamics of
inhibitory networks of spiking neurons with arbitrarily complicated
connectivity and delayed interactions. We showed that all generic
trajectories are stable, even if they generate highly irregular dynamics.
In particular, the results analytically confirm recent numerical findings
\cite{Zillmer2006} that irregular dynamics in spiking neural networks
may exhibit  stable behavior. Moreover, as for globally coupled systems
without delay \cite{Jin2002} our results show that also networks
with more complex structure and delayed interactions exhibit stable
periodic orbits. However, highly irregular yet stable transient trajectories
dominate the dynamics in all large and non-globally coupled networks,
in stark contrast to the fast convergence to attractor dynamics found
recently in globally coupled networks \cite{Jin2002} and also opposite
to the long chaotic (and thus unstable) transients in randomly diluted
networks of excitatory neurons \cite{Zumdieck2004}. In particular,
in the class of systems presented here, almost every trajectory 
is stable, not only if it generates regular \cite{Timme2002} but
even if it generates very irregular dynamics. Curiously, the assumption
of events generated by a maximally irregular (Poisson) random process
led us to show stability of the deterministic trajectories. 

More generally, our results underline that multi-dimensional deterministic
dynamics with statistical properties close to that of a random system
need not be generated by deterministic chaos. As we have shown analytically
for spiking neural networks with delayed interactions, dynamical irregularity
may well be generated by stable trajectories, dominantly on a transient
not on an attractor. Future work needs to investigate closer the key
consequences for systems in which irregular dynamics is stable. For
instance stable irregular dynamics lifts the important practical constraint
of long-term unpredictability that irregular dynamics bears if it
is generated by chaos. Stable irregular dynamics, even in multi-dimensional
systems, may well be predictable in practice. For the networks of
spiking neurons studied above, this has the astounding consequence
that  the dynamics in only a small time window, even in the presence
of some errors, defines the entire future of the highly irregular
spiking dynamics.

We thank Fred Wolf for helpful comments and the Federal Ministry of
Education and Research (BMBF) Germany for partial support under Grant
No. 01GQ0430.


\begin{thebibliography}{19}
\bibitem[{Del()}]{Delay}
\bibinfo{note}{Delays make the system formally infinite dimensional; for the
  systems considered here, however, the dimension is generally high but finite
  \cite{Ashwin2005}.}
\end{thebibliography}
\end{document}